# Simple Generator of Variable Test Tasks Compatible IMS QTI

A. D. Zaikin

Novosibirsk State Technical University, Novosibirsk, Russia

*The author details a method of generating variable test tasks compliant with the IMS QTI specification, which includes preparation of tasks in an electronic table with the further formation of XML files using VBA macros. The following types of questions are supported: Single Choice, Multiple Choice, and Numerical. Implementing the method requires only primary programming skills. The ease of implementation and flexibility of the proposed generation method provides ample opportunities for testing at all stages of the learning process. The method was tested during a physics course for engineering students at a technical university.*

Computer-based testing systems have become an integral part of the learning process, and are extensively used at every stage of teacher-student interaction in distance and class-based learning alike. Computer-based testing allows students to control the level of their knowledge on their own and facilitates both monitoring and teaching functions. A teacher can interact with a group of students and minimize the time spent checking knowledge acquisition.

Modern wide-spread distance learning systems, such as Web Tutor, Lotus, Learning Space, Prometheus, and Moodle, include a testing module, and many higher education institutions have their own distance learning systems. Thus, the Institute of Distance Education (IDE) at Novosibirsk State Technical University has developed and uses DiSpace 2.0 [1, 2], a distance learning system that includes the DiTest v2testing sub-system.

DiTest v2 is a remote testing system [3] that supports the international Question&Test Interoperability (QTI) specification (version 2.0) from IMS [4]. This solution helps standardize learning technologies, unify test task databases used in various software-based testing systems, make databases independent of the program shell and solve the problem of exchanging test task banks.

The system is accessed using a browser, making it possible to conduct testing in field conditions. The terminal class is well suited to substitute student smartphones.

DiTest v2 consists of three autonomous sub-systems (Test Entry/Editing, Survey, and Results Processing and Statistics Gathering), and two databases (Test Tasks Database and Test Surveys Database).

The Test Entry/Editing sub-system is implemented as a graphic shell and contains a mathematical formula editor based on the open MathML specification. Along with mathematical formulas, tests may also include media and metadata. The sub-system outputs an XML text file complaint with the IMS QTI specification.

Unfortunately, the reality is that unscrupulous students will always look for easy ways to pass tests. Thus, if students are going to take a static test, they often try to remember the numbers of correct answers instead of studying the subject matter. It is no secret that in this case every next group performs better on the test than the previous one. A simple internet search for exam answers, term papers or tests paints a sad picture. The grey market for pseudo-educational services is extremely developed.

In our opinion, one solution to this challenge is the creation of individual, variable and mobile tests. Teachers should have the ability to promptly change tasks (within reasonable limits, of course), and prepare them for a certain class in a certain group.

However, the test creation procedure in DiTest v2 and similar systems cannot be called simple. The price of joining for teachers is rather high, including not only subject-related and methodological issues, but also understanding the operation principles of a certain testing system and mastering the graphic shell and program modules. Even after mastering the testing system, the labor involved in entering a test originally composed on paper is rather high, which is especially relevant for teachers not directly related to computer science.

The author goes on to provide a set of solutions developed and used in the process of teaching physics for engineering disciplines at a technical university. The nature of this subject makes it possible to abandon many of the more "exotic" capabilities of the QTI specification. Using multimedia, placing objects on a map, using adaptive units that determine the system's dynamic behavior, in correct user answers, and many other features are beyond the scope of our research. The algorithm of a student taking a test is the following: a question is asked, an answer is given, and an assessment is made. Such restrictions make it possible to implement a simple and convenient algorithm of test development.

DiTest v2 supports the following intuitive types of questions: Single Choice, Multiple Choice, Numerical, Associations, Matching, Short Answer, Sorting, and Open-Ended.

The first three are most relevant for the stated objectives. The Single Choice type involves choosing one of several proposed answers. The answer will count if the student chooses it correctly. The Multiple Choice type means that several answers can be chosen. These types of questions can be used to check special knowledge or the knowledge of instructional material for a specific student. The Numerical type of question requires an answer in the form of a number obtained as the result of calculations. In fact, it is a task, not a test question. Students are expected to know physical laws and formulas, apply them, and complete calculations to find a finite number.

As an example of a Numerical question, let's consider the following task from the Mechanics section.

*Two material points collide. The mass of the first material point is **two** times **more than** the mass of the second point. The collision is fully inelastic. The speed of the first point before the collision is $V_x$=**-2** m/s and $V_y$=**7** m/s, and the speed of the second point is $V_x$=**7** m/s and $V_y$=**4** m/s. Find the speed module of the compound particle. Give the answer in **decimeters per second** rounded to the nearest whole number.*

It is not difficult to create several versions of this task. First of all, we can change the numerical values of the speed component. But there is even more variability if we change not only the numbers, but also the words in bold in the task condition. For example, according to the following pattern: **two → one and a half→ three→…; more than→ less than; decimeters → centimeters → millimeters**. In this case, the test author will have to calculate the answer for each set of parameters and expound it in the required form.

Entering such questions in the Test Entry/Editing sub-system, even by copy-pasting, requires a great deal of manual work that might entail errors, along with other issues, which always take place in the event of manual data entry.

It should be noted that the task text contains both an invariable and variable part. In our opinion, they can be combined using a popular and generally accessible software product such as Excel.

Let's give each test question one Excel line, the cells of which are filled in according to certain markup rules. Each of the variable parameters and the exact numerical answer are placed in individual cells. The question text is formed using Excel functions combining the invariable and variable parts of the question.

Let's place the word "two" in the *D1* cell and the words "more than" in the *C1* cell, then in the cell containing the formula

=*"The mass of the first material point is '& D1 &' times '& C1 &' the mass of the second point."*

there will be the following text

*The mass of the first material point is **two** times **more than** the mass of the second point.*

Let's fill in columns D and C in random order with possible variants of variable parameters, and then by dragging the filling marker over the column containing the question text, we will get many variants of the above sentence. Of course, this method applies to all variable parameters of the problem.

We can also easily fill in the column with exact numerical answers. To do so, it is enough to write the applicable formula in one cell and drag the fill marker. For the given task, the answer, if $k$ is the ratio of masses of the first and second bodies, takes the following form:

$$V_3 = \sqrt{V_{3x}{}^2 + V_{3y}{}^2} = \sqrt{\left(kV_{1x} + V_{2x}\right)^2 + \left(kV_{1y} + V_{2y}\right)^2} \Big/ (k+1)$$

Thus, we can receive a set of tasks with exact numerical answers, each of which contains individual numerical parameters.

In some tasks, the choice of variable parameters is complicated due to various restrictions. Let's consider, for example, the following task:

*A material point is moving based on the equation: $x=A+Bt+Ct^2+Dt^3$, where A=-5.0 m, B=2.0 m/s, C=2.0 m/s², D=-1.0 m/s³. Find the coordinate of the material point at the moment its speed equals zero.* The material point will stop only if the coefficients have the ratio:

$$C^2 - 3BD > 0 , \quad BD < 0 .$$

Such restrictions can also be checked in an electronic table. By writing the applicable formula in the cell and dragging the fill marker, we can identify and reject the variable parameters that do not meet the established conditions.

Formulating a group of questions is only the first part of the issue. The second part is transferring this group into the testing system. Importing tests into the testing system in the QTI specification is mandatory, and DiTest v2 supports this, of course. To implement this task, a set of VBA macros is written. Macro commands (or macros) refer to the sequences of commands. VBA macro is a macro written in Visual Basic for Applications supported by Microsoft Office, including Excel.

As mentioned above, according to the QTI specification, each test question is a separate XML text file. The structure of an XML file can be presented as an element tree. Every element starts with an opening tag in angle brackets, then comes the element content, and after that comes the closing tag in angle brackets with a slash (/) closing the element. An example of a file with a numerical question is provided below. The file structure is intentionally simplified for illustrative purposes. Forming such a file within VBA requires only basic programming skills.

```xml
<?xml version="1.0" encoding="UTF-8"?>
<assessmentItemtitle="Question1" identifier="Item_1.....>
        <responseDeclarationidentifier="RESPONSE" baseType="string" cardinality="single">
                <correctResponse><value>61</value></correctResponse>
                <mappingdefaultValue="0" upperBound="1" lowerBound="0"/>
        </responseDeclaration>
        <itemBody>
                <blockquote>
                <![CDATA[<p>Two material points collide. The mass of the first material point is two times more than
                the mass of the second point. The collision is fully inelastic. The speed of the first point before the
                collision is <em>V<sub>x</sub> =-2 m/s, V<sub>y</sub>=7 m/s</em>, and the speed of the second
                point is <em>V<sub>x</sub> =7 m/s, V<sub>y</sub>=4 m/s</em>. <p> Find the speed module of the
                compound particle. Give the answer in decimeters per second rounded to the nearest whole
                number.</p></p>]]>
                </blockquote>
        </itemBody>
</assessmentItem>
```

The minimum required differences of one test question from another are highlighted in bold. They include:

− the *title* and *identifier* attributes of the core element *assessmentItem;*

− the *value* child element of the element *correctResponse;*

    − the child element *blockquote* of the element *itemBody*.

The *title* attribute is the question heading visible only to the test developer. The *identifier* attribute is a unique identifier of the test question, which also determines the file name (*Item_1.xml* in this case). The *value* element contains the correct numerical answer, which is 61 dm/s in this task.

The *blockquote* element contains the question text. The hypertext markup language of HTML documents using special tags to define the formatting of the document contents makes it possible to write simple formulas directly in the body of the question placed in the XML file. In the given example, it is shown how to use paragraph, italics, upper and lower indices.

More complicated formatting can be created based on the open MathML specification. However, this makes the test structure bulky and less flexible. Such expanded capabilities are beyond the objective set.

Besides the above task condition, the question in the test is followed by certain qualifying information, such as didactic unit, question subject, question number, question type, maximum points for the question, recommended test time, comments to the answer, etc. Part of these optional parameters can be omitted, and part is determined by default. Parameters that change from question to question or from test to test are placed in the cells of the question line, and are taken into account when forming an XML file. Thus, in the given example, the *upperBound* attribute of the *<mapping>* tag determines that the maximum points for the correct answer is "1". If required, the value of this attribute can be changed for any other value.

A test is a group of questions with a related set of rules that determine which elements and in what order the user will see and interact with them. The group of test questions is in fact a set of XML text files. They are generally supplemented with hypermedia content in the HTML/XHTML format.

Thus, we need a closing stage: the formation of an XML file describing the general structure and interrelation of the interactive elements (questions) of the test. It should be noted that all files are then placed in a ZIP container.

A sample of the file structure simplified for illustration purposes is provided below. The test consists of two questions. This file can also be easily formed using VBA.

```xml
<?xml version="1.0" encoding="UTF-8"?>
<assessmentTest title="Mechanics" identifier="Test_1"...>
        <testPart identifier="Part_1" submissionMode="simultaneous" navigationMode="linear">
        <assessmentSection title="Momentum Conservation Principle " identifier="Section_1"...>
                <assessmentItemRef identifier="Item_1" href="item_1.xml">
                        <itemSessionControl showSolution="true"/>
                </assessmentItemRef>
                <assessmentItemRef identifier="Item_2" href="item_2.xml">
                        <itemSessionControl showSolution="true"/>
                </assessmentItemRef>
        </assessmentSection>
        </testPart>
</assessmentTest>
```

The mandatory attributes *Title* and *Identifier* of the *<assessmentTest>* tag determine the short name and unique identifier of the test in the system. The *<testPart>* tag determines the section of an upper level (didactic unit), which can also include several sections (subjects), the tag is *<assessmentSection>*. These tags allow the author to structure their test. The *identifier* and *href* attributes of the *<assessmentItemRef>* tag determine the question included in the subject and its corresponding file.

The method used to create a test containing Single Choice and Multiple Choice types of questions does not have any principle differences from what is described above. Each test question has a corresponding separate line in the electronic table. The XML files comprising the test are formed using VBA macros. Of course, the structure of the tag tree of the file that contains the question will in this case be different, while the file describing the test will generally remain the same as described above.

The test containing the task (numerical question) with individual parameters can be easily formed both for an academic group, including about 30 students, and for a flow consisting of five groups of about 150 students. Each student is supposed to answer only one test question in accordance with the pattern determined by the teacher. Of course, not every task can be included in this kind of test. Difficult, olympiad, or other tasks that require more calculations or the use of charts, diagrams or pictures are better used in another field.

Such tests are intended to form and check the primary skills of students, including beyond the subject area. Yesterday's graduates often have difficulties performing elementary operations, including exact calculations, rounding, interest calculation, and the use of multiple and sub-multiple units. The easy creation of multiple choice tests turns them into a weapon of mass destruction for teachers. Every student is reached, and checking the correctness of calculations is the algorithm's responsibility. Tests take just 5 to 10 minutes in practically any class, as students are inseparable from their smartphones.

Several types of such tests are proposed. Despite time constraints, students can be objectively scaled by knowledge and skills. Another option is to form an individual test for each student containing several tasks (questions), with all tasks taken from a single bank with various numeric parameters. This option is good for review exercises.

Single Choice and Multiple Choice questions are optimal for checking student knowledge of terminology and laws. Another option for such tests is using them as a simulator for repeated individual work to help students learn the material. Illustrative questions and their convenient manipulation make it possible to create a test immediately before a class in accordance with a specific learning situation. For example, by preparing a short test of 2-3 questions on the lecture subject, teachers can combine a short test of student attentiveness and knowledge, and the electronic registration of class attendance.

# ПРОСТОЙ ГЕНЕРАТОР ВАРИАТИВНЫХ ТЕСТОВЫХ ЗАДАНИЙ, СОВМЕСТИМЫХ С IMS QTI

А.Д. Заикин

Новосибирский государственный технический университет,
г. Новосибирск, Россия

*Предложен способ генерации вариативных тестовых заданий, соответствующих спецификации IMS QTI, включающий в себя подготовку заданий в электронной таблице с последующим формированием файлов формата XML посредством VBA макросов. Поддерживаются следующие типы вопросов: «Одиночный», «Множественный», «Числовой», «Короткий». Для реализации метода достаточно начальных навыков программирования. Простота реализации, гибкость предложенного метода генерации открывают широкие возможности для применения тестирования на всех стадиях учебного процесса. Методика апробирована в процессе преподавания курса физики студентам инженерных специальностей технического университета.*

*Ключевые слова: тестирование, система DiTest v2, спецификация IMS QTI, макрос VBA, тренажер.*

Системы компьютерного тестирования стали неотъемлемой частью образовательного процесса, находя применение на всех стадиях взаимодействия преподаватель-студент, как в системах дистанционного, так и очного обучения. Компьютерное тестирование, позволяя студенту самостоятельно контролировать уровень знаний, обладает при этом не только контролирующими, но и обучающими функциями. Преподаватель же получает возможность взаимодействовать с группой студентов, минимизировать время на проверку знаний.

Современные системы дистанционного обучения, получившие широкое распространение, такие как WebTutor, Lotus LearningSpace, Прометей, Moodle, имеют подсистему тестирования. Многие вузы разрабатывают свои системы дистанционного обучения. Так в НГТУ Институтом дистанционного образования (ИДО) разработана и используется система дистанционного обучения DiSpace 2.0 [1,2], включающая подсистему тестирования DiTest v2.

DiTest v2 – система удаленного тестирования [3], поддерживающая международную спецификацию Question&Test Interoperability (QTI) версии 2.0 консорциума IMS [4]. Данное решение позволяет стандартизировать обучающие технологий, унифицировать базы данных тестовых заданий, используемых в различных программных тестирующих системах, сделать базы данных независимыми от программной оболочки, решить проблему обмена банками тестовых заданий.

Инструментом доступа к системе служит браузер, что позволяет проводить тестирование не только в терминальных классах, оснащенных компьютерами, но и в обычных аудиториях. В этом случае студент взаимодействует с системой посредством смартфона.

Система DiTest v2 состоит из трех автономных подсистем: «Ввода/редактирования тестов», «Проведение опроса», «Обработки результатов и сбора статистики», и двух баз данных: «Базы данных тестовых заданий, и «Базы данных тестовых опросов».

Подсистема «Ввода/редактирования тестов», реализованная в виде графической оболочки, содержит редактор математических формул, базирующийся на открытой спецификации MathML. Наряду с математическими формулами в тесте можно использовать медиа и метаданные. На выходе подсистемы – текстовый файл в формате XML, удовлетворяющий спецификации IMS QTI.

Широкое использование систем тестирования в учебном процессе выявляет ограничения и узкие места таких методик. Так, работая со статичным тестом, студент нередко не предмет осваивает, а запоминает номера правильных ответов. Как правило, каждая последующая группа выполняет тест лучше, чем предыдущая.

Одним из ответов на этот вызов, на наш взгляд, является сознание индивидуальных, вариативных, мобильных тестов. Преподаватель должен иметь возможность оперативно изменять задания, в разумных пределах, конечно, готовить их конкретному занятию в конкретной группе.

Однако процедуру создания тестов как в системе DiTest v2, так и в аналогичных системах, нельзя назвать простой. Порог вхождения для преподавателя весьма высок. Ему необходимо решить не только собственно предметно-методологические проблемы, но и понять принципы работы конкретной системы тестирования, освоить оболочку и программные модули. Даже после освоения системы тестирования трудоемкость создания в ней теста от идеи до конечного продукта достаточно велика. Особенно это актуально для преподавателей, непосредственно не связанных с информатикой.

Далее излагается набор решений, разработанных и использованных в процессе изучения дисциплины физики на инженерных специальностях технического университета. Особенности предмета позволяют отказаться от многих экзотических возможностей спецификации QTI. Использование мультимедиа, размещение объектов на карте, адаптивных блоков, определяющих динамическое поведение системы, когда пользователь дает неправильный ответ, и многое другое оставим за рамками наших построений. Алгоритм работы студента с тестом предполагается следующий: вопрос поставлен, ответ дан, оценка определена. Подобные ограничения позволили реализовать простой и удобный алгоритм разработки теста.

Система DiTest v2 поддерживает следующие интуитивно понятные типы вопросов: «Одиночный», «Множественный», «Короткий», «Числовой», «Ассоциации», «Соответствие», «Упорядочение», «Открытый».

Для реализации поставленных целей актуальны первые четыре. Тип вопроса «Одиночный» предполагает выбор одного варианта ответа из нескольких предложенных. Ответ засчитывается, если студент выбрал его правильно. Тип вопроса «Множественный» предполагает, что можно выбрать несколько вариантов ответа. Тип вопроса «Короткий» предполагает, что студент, не имея возможности выбрать правильный ответ, должен ввести его в текстовое поле ввода. Эти типы вопросов позволяют проверить специальные знания, знания учебного материала у конкретного студента.

Тип вопроса «Числовой» предполагает указание ответа в форме числа, которое получается в процессе расчета. По сути это задача, а не вопрос теста. Предполагается, что студент знает физические законы и формулы, может их применить, довести расчеты до конечного числа.

В качестве примера типа вопроса «Числовой» рассмотрим следующую задачу из раздела механика.

*Сталкиваются две материальные точки. Масса первой материальной точки в **два раза больше** массы второй. Удар абсолютно неупругий. Скорость первой точки перед столкновением $V_x$=-2 м/с, $V_y$=7 м/с, скорость второй – $V_x$=7 м/с, $V_y$=4 м/с. Найти модуль скорость составной частицы. Ответ, округлив до целых, дать в **дециметрах** в секунду.*

Создать несколько вариантов этой задачи нетрудно. Прежде всего, конечно, можно менять числовые значения компонент скоростей. Еще большая вариативность возникает при заменах в условии задачи не только чисел, но и выделенных слов. Например, по следующей схеме: ***два → полтора → три →…; больше → меньше; дециметрах → сантиметрах → миллиметрах***. При этом составителю теста потребуется для каждого комплекта параметров вычислить ответ и привести его к требуемой форме.

Ввод таких вопросов в подсистему «Ввода/редактирования тестов», даже при использовании функций копировать-вставить, требует большого объема ручной работы, которая кроме всего чревата ошибками, всегда возникающими при ручном вводе информации.

Заметим, что текст задачи содержит неизменную и вариативную части. Объединять их, на наш взгляд, лучше всего используя такой популярный и общедоступный программный продукт как Excel.

Каждому вопросу теста поставим в соответствие одну строчку таблицы Excel, ячейки которой заполняются в соответствие с некоторыми правилами разметки. Каждый из варьируемых параметров и точный числовой ответ размещаются в отдельных ячейках. Текст вопроса формируется текстовыми функциями Excel, с помощью которых объединяются неизменная и вариативная часть вопроса.

Покажем это на примере двух вариативных параметрах. Пусть в ячейке *D1* размещается слово *"два"*, а в ячейке *C1* – *"больше"*, тогда в ячейке *B1,* содержащей формулу

   ="Масса первой материальной точки в " & D1 & " раза " & C1 & " массы второй."

будет находиться текст

*Масса первой материальной точки в **два** раза **больше** массы второй.*

Заполним в случайном порядке колонки *D* и *C* возможными вариантами вариативных параметров, тогда перетаскивая маркер заполнения по колонке, содержащей текст вопроса, получим много вариантов приведенного выше предложения. Разумеется, этот метод применяется ко всем вариативным параметрам задачи.

Также легко заполняется колонка с точными числовыми ответами. Для этого достаточно записать соответствующую формулу в одной ячейке и перетащить маркер заполнения. Для приведенной задачи ответ, если *k* – отношение масс первого и второго тел, формула имеет вид

$$V_3 = \sqrt{V_{3x}^{\ 2} + V_{3y}^{\ 2}} = \sqrt{\left(kV_{1x} + V_{2x}\right)^2 + \left(kV_{1y} + V_{2y}\right)^2} \Big/ \left(k+1\right)$$

Таким образом, можно получить набор задач с точными числовыми ответами, каждая из которых содержит индивидуальные числовые параметры.

В некоторых задачах выбор вариативных параметров осложнен разнообразными ограничениями. Для примера рассмотрим приведенную ниже задачу.

*Материальная точка движется согласно уравнению x=A+Bt+Ct²+Dt³, где A=-5.0 м, B=2.0 м/с, C=2.0 м/с², D=-1.0 м/с³. Найти координату материальной точки в тот момент, когда ее скорость станет равной нулю.*

Задача имеет решение, только если между коэффициентами выполняется соотношение

$$C^2 - 3BD > 0 \ , \quad BD < 0 \ .$$

В противном случае материальная точка не остановится.

Проверку такого рода ограничений также удобно проводить в электронной таблице. Записав соответствующую формулу в ячейку и перетащив маркер заполнения, можно выявить и отбраковать вариативные параметры, не удовлетворяющие заданным условиям.

Построение массива вопросов – это только первая часть проблемы. Вторая часть – перенос этого массива в систему тестирования. Возможность импорта тестов в тестирующую систему по спецификации QTI является обязательной, и система DiTest v2, конечно же, ее поддерживает. Для реализации этой задачи написан набор макросов VBA. Макрокомандами (сокращенно – макросами) называются последовательности команд. Макрос VBA – это макрос, написанный на языке Visual Basic for Applications, поддерживаемый приложениями Microsoft Office, в том числе и Excel.

Как уже говорилось выше, в соответствии со спецификацией QTI каждый вопрос теста – это отдельный текстовый файл в формате XML. Структура XML файла может быть представлена в виде дерева элементов. Элемент начинается открывающим тегом в угловых скобках, затем идет содержимое элемента, после него – закрывающий тег в угловых скобках, с символом слеш / завершения элемента. Пример файла, содержащего численный вопрос, приведен ниже. Структура файла намеренно упрощена для наглядности. Для формирования такого файла в рамках VBA достаточно самых начальных навыков программирования.

```xml
<?xml version="1.0" encoding="UTF-8"?>
<assessmentItem title="Вопрос1" identifier="Item_1".....>
        <responseDeclaration identifier="RESPONSE" baseType="string" cardinality="single">
                <correctResponse><value>61</value></correctResponse>
                <mapping defaultValue="0" upperBound="1" lowerBound="0"/>
        </responseDeclaration>
        <itemBody>
                <blockquote>
                <![CDATA[ <p>Сталкиваются две материальные точки. Масса первой в два раза больше массы
                второй. Удар абсолютно неупругий. Скорость первой точки перед столкновением
                <em>V<sub>x</sub> =-2 м/с, V<sub>y</sub>=7 м/с</em>, второй - <em>V<sub>x</sub> =7 м/с,
                V<sub>y</sub>=4 м/с</em>. <p>Найти модуль скорость составной частицы. Ответ, округлив до
                целого, дать в дециметрах в секунду.</p> </p> ]]>
                </blockquote>
        </itemBody>
</assessmentItem>
```

Выделенный текст – это содержимое элементов, изменяющееся в различных вопросах теста. Вариативные элементы включают:

− атрибуты *title* и *identifier* корневого элемента *assessmentItem*;

− дочерний элемент *value* элемента *correctResponse*;

− дочерний элемент *blockquote* элемента *itemBody*.

Атрибут *title* – заголовок вопроса, видимый только разработчику теста. Атрибут *identifier* – уникальный идентификатор вопроса в тесте, определяющий также имя файла, *Item_1.xml* в данном случае. Элемент *value* содержит правильный числовой ответ равный в данной задаче 61 дм/с.

Элемент *blockquote* содержит текст вопроса. Язык гипертекстовой разметки документов HTML, использующий специальные теги для обозначения форматирования содержимого документов, позволяет простые формулы записывать непосредственно в теле вопроса, помещаемого в XML файле. В приведенном примере показано, как применяя, соответствующие теги, использовать в тексте вопроса курсив, абзац, верхние и нижние индексы.

Более сложное форматирование можно создавать на основе открытой спецификации MathML, однако это делает конструкцию теста громоздкой и менее гибкой. Подобное расширение возможностей лежит за рамками поставленной задачи.

Кроме приведенного выше условия задачи вопрос в тесте сопровождает некоторая служебная информация, как то: дидактическая единица, тема вопроса, номер вопроса, тип вопроса, максимальная оценка за вопрос, рекомендуемое время прохождения теста, комментарий к решению и т.п. Часть из этих параметров, являющихся необязательными, может быть опущена, часть определяться по умолчанию. Параметры, изменяющиеся от вопроса к вопросу или от теста к тесту, размещаются в ячейках строки вопроса, и учитываются при формировании XML файла. Так в приведенном примере атрибут *upperBound* тега *<mapping>* определяет, что максимальная оценка за правильный ответ равняется "1". При желании значение этого атрибута изменяется на любое другое.

Тест – это массив вопросов со связанным набором правил, которые определяют, какие элементы и в каком порядке видит пользователь, и каким образом он взаимодействует с ними. Массив же вопросов теста есть, по сути, набор текстовых файлов в формате XML. В общем случае к ним добавляется еще гипермедийный контент в формате HTML/XHTML.

Таким образом, необходим завершающий этап – формирование XML файл, описывающего общую структуру и взаимосвязь интерактивных элементов (вопросов) теста. Заметим, что все файлы затем помещаются в ZIP-контейнер.

Образец файла с упрощенной для наглядности структурой приведен ниже. Тест состоит из двух вопросов. Формирование такого файла средствами VBA также не вызывает трудностей.

```xml
<?xml version="1.0" encoding="UTF-8"?>
<assessmentTest title="Механика" identifier="Test_1"...>
        <testPart identifier="Part_1" submissionMode="simultaneous" navigationMode="linear">
        <assessmentSection title="Закон сохранения импульса" identifier="Section_1"...>
                <assessmentItemRef identifier="Item_1" href="item_1.xml">
                        <itemSessionControl showSolution="true"/>
                </assessmentItemRef>
                <assessmentItemRef identifier="Item_2" href="item_2.xml">
                        <itemSessionControl showSolution="true"/>
                </assessmentItemRef>
        </assessmentSection>
        </testPart>
</assessmentTest>
```

Обязательные атрибуты *title* и *identifier* тега *<assessmentTest>* определяют краткое название и уникальный идентификатор теста в системе. Тег *<testPart>* определяет раздел верхнего уровня (дидактическую единицу), которая в свою очередь может содержать несколько секций (тем), тег – *<assessmentSection>*. Эти теги позволяют автору структурировать тест. Атрибуты *identifier* и *href* тега *<assessmentItemRef>* определяют вопрос, включенный в тему, и соответствующий ему файл.

Методология формирования теста, содержащего вопросы «Одиночный», «Множественный» и «Короткий», не имеет принципиальных отличий от описанной выше. Каждому вопросу теста соответствует отдельная строка в электронной таблице. Файлы формата XML, составляющие тест, формируются средствами макросов VBA. Конечно, структура дерева тегов файла, содержащего вопрос, в этом случае будет иная, а вот файл, описывающий тест в целом останется таким же, как и приведенный выше.

Тест, содержащий задачу (числовой вопрос) с индивидуальными параметрами, легко сформировать как для академической группы, включающей ~30 студентов, так и для потока, состоящего из пяти групп, ~150 студентов. Конкретному студенту предназначен лишь один вопрос теста в соответствии с определенной преподавателем схемой. Разумеется, не любая задача годится для включения в такой тест. Сложным, олимпиадным задачам, задачам, требующим большого объема вычислений или присутствия в условии схем, графиков и рисунков, лучше найти иную область применения.

Такого рода тесты предназначены для формирования и проверки наличия у студента простейших навыков и не только в предметной области. Вчерашний абитуриент зачастую испытывает затруднения при выполнении элементарных операций. Точные вычисления, округление, расчет процентов, использование кратных и дольных единиц – все это является камнем преткновения для многих. Простота создания вариативных тестов превращает их в эффективный инструмент в руках преподавателя. Задание носит персонифицированный характер, а проверка правильности расчетов возложена на алгоритм. Тесту можно уделить буквально несколько минут практически на любом занятии, благо смартфон стал неотъемлемым атрибутом современного студента.

Предлагая несколько таких тестов в условиях ограниченного лимита времени, можно объективно ранжировать студентов по знаниям и умениям. Другой вариант использования – формировать для каждого студента индивидуальный тест, содержащий несколько задач (вопросов), разумеется, все задачи берутся из единого банка с различающимися числовыми параметрами. Этот вариант подходит для проведения контрольной работы.

Вопросы «Одиночный», «Множественный» и «Короткий» оптимальны для проверки знаний терминологии и законов. Один из вариантов применения таких тестов – использовать их как тренажер, который предполагает самостоятельное и многократное прохождение задания с целью помочь студенту закрепить материал. Наглядность представления вопросов, удобство манипулирования ими позволяют формировать тест непосредственно перед занятием, в соответствии с конкретной учебной ситуацией. Например, подготовив перед лекцией короткий тест из 2-3 вопросов по теме лекции, можно совместить тест на внимательность студента и усвояемость материала, и электронную регистрацию присутствующих.

## ЛИТЕРАТУРА